\documentclass[conference]{IEEEtran}
\IEEEoverridecommandlockouts
\usepackage{cite}
\usepackage{amsmath,amssymb,amsfonts}
\usepackage{algorithmic}
\usepackage{graphicx}
\usepackage{textcomp}
\usepackage{xcolor}
\usepackage{algorithm}
\usepackage{enumerate}

\newcommand{\tabincell}[2]{\begin{tabular}{@{}#1@{}}#2\end{tabular}} 
\usepackage{array}
\usepackage[caption=false,font=normalsize,labelfont=sf,textfont=sf]{subfig}
\usepackage{textcomp}
\usepackage{stfloats}
\usepackage{url}
\usepackage{verbatim}
\usepackage{cite}
\usepackage{makecell}
\usepackage{multirow}
\usepackage{booktabs}
\usepackage{footnote}
\usepackage{threeparttable}
\usepackage{epstopdf}
\usepackage{fancyhdr}

\def\BibTeX{{\rm B\kern-.05em{\sc i\kern-.025em b}\kern-.08em
    T\kern-.1667em\lower.7ex\hbox{E}\kern-.125emX}}
\begin{document}
\title{SpOctA: A 3D Sparse Convolution Accelerator with Octree-Encoding-Based Map Search and Inherent Sparsity-Aware Processing}
\author{\IEEEauthorblockN{Dongxu Lyu$^{1}$, Zhenyu Li$^{1}$, Yuzhou Chen$^{1}$, Jinming Zhang$^{1}$, Ningyi Xu$^{1,3}$, Guanghui He$^{1,2,3*}$}
\IEEEauthorblockA{\textit{$^{1}$School of Electronic Information and Electrical Engineering, Shanghai Jiao Tong University, Shanghai, China} \\
\textit{$^2$MoE Key Lab of Artificial Intelligence, AI Institute, Shanghai Jiao Tong University, China}
\textit{$^{3}$Huixi Technology}\\
\{lvdongxu, ambitious-lzy, huygens, tiaozhanzhe, xuningyi, guanghui.he\}@sjtu.edu.cn
}
}
\maketitle
\begin{abstract}
Point-cloud-based 3D perception has attracted great attention in various applications including robotics, autonomous driving and AR/VR.
In particular, the 3D sparse convolution (SpConv) network has emerged as one of the most popular backbones due to its excellent performance.
However, it poses severe challenges to real-time perception on general-purpose platforms, such as lengthy map search latency, high computation cost, and enormous memory footprint.
In this paper, we propose SpOctA, a SpConv accelerator that enables high-speed and energy-efficient point cloud processing.
SpOctA parallelizes the map search by utilizing algorithm-architecture co-optimization based on octree encoding, thereby achieving 8.8-21.2$\times$ search speedup.
It also attenuates the heavy computational workload by exploiting inherent sparsity of each voxel, which eliminates computation redundancy and saves 44.4-79.1\% processing latency.
To optimize on-chip memory management, a SpConv-oriented non-uniform caching strategy is introduced to reduce external memory access energy by 57.6\% on average. 
Implemented on a 40nm technology and extensively evaluated on representative benchmarks, SpOctA rivals the state-of-the-art SpConv accelerators by 1.1-6.9$\times$ speedup with 1.5-3.1$\times$ energy efficiency improvement.

\end{abstract}


\section{Introduction} \label{sec:I_Introduction}

Point clouds, a collection of points obtained from a variety of 3D sensors such as LiDAR scanners and RGB-D cameras, contain extensive geometric, shape and depth information\cite{PCSurvey_PAMI21}.
In contrast to 2D images, point clouds provide an exceptional opportunity for a reliable and robust understanding of the surrounding environment \cite{VoxelNet_CVPR18}.
Consequently, point clouds have become an indispensable modality in various promising applications, including virtual reality (VR), augmented reality (AR), robotics and autonomous driving (AD).
These practical applications explicitly impose stringent necessities on network processing, requiring low latency, high accuracy and acceptable energy consumption \cite{PCSurvey_PAMI21}.

Unlike images, 3D point clouds exhibit randomly spatial sparsity, thus necessitating distinct processing pattern from 2D convolutional neural networks (CNN)\cite{ResNet_CVPR16}.
To address the sparsity and irregularity of point clouds, researchers commonly choose to voxelize them into a uniform tensor representation \cite{VoxelNet_CVPR18} and introduce 3D sparse convolution (SpConv)\cite{Subm_CVPR19} as the network backbone\cite{MinkUNet_CVPR19,SECOND_Sensors18,CenterPoint_CVPR21,NASMIT_ECCV20}.
While SpConv-based models deliver satisfactory accuracy, they often run at under 20 frames per second (fps) on powerful NVIDIA GPUs\cite{CenterPoint_CVPR21} due to the inefficient map search and high computation cost.
In comparison, ResNet \cite{ResNet_CVPR16} runs at more than 800 fps.
The high latency and unacceptable power consumption constrain the applicability of 3D SpConv models in common industrial solutions \cite{TorchSparse_MLSys22}.

The new operator, map search in SpConv, causes extreme memory inefficiency and large processing latency explicitly.
SpConv requires serial traversing for each voxel's neighbors to build the IN-OUT maps for matrix multiplication, leading to frequent and irregular external memory access. 
Although hash-table-based approach is commonly employed in GPU-based optimizations \cite{TorchSparse_MLSys22,MinkUNet_CVPR19,spconv2022} to simplify search access, it consumes excessive overhead for edge devices and cannot be parallelized\cite{PointAcc_MICRO21}.
The state-of-the-art SpConv accelerators also remain stuck in the inevitable serial processing \cite{PointAcc_MICRO21,SpConvASIC_VLSI22} or heavy logic consumption for parallel search circuits\cite{Tsinghua_ISSCC23}, confined to restricted acceleration and unsatisfactory logic simplification.

\begin{figure}[t]
    \centering
    \includegraphics[width=\linewidth]{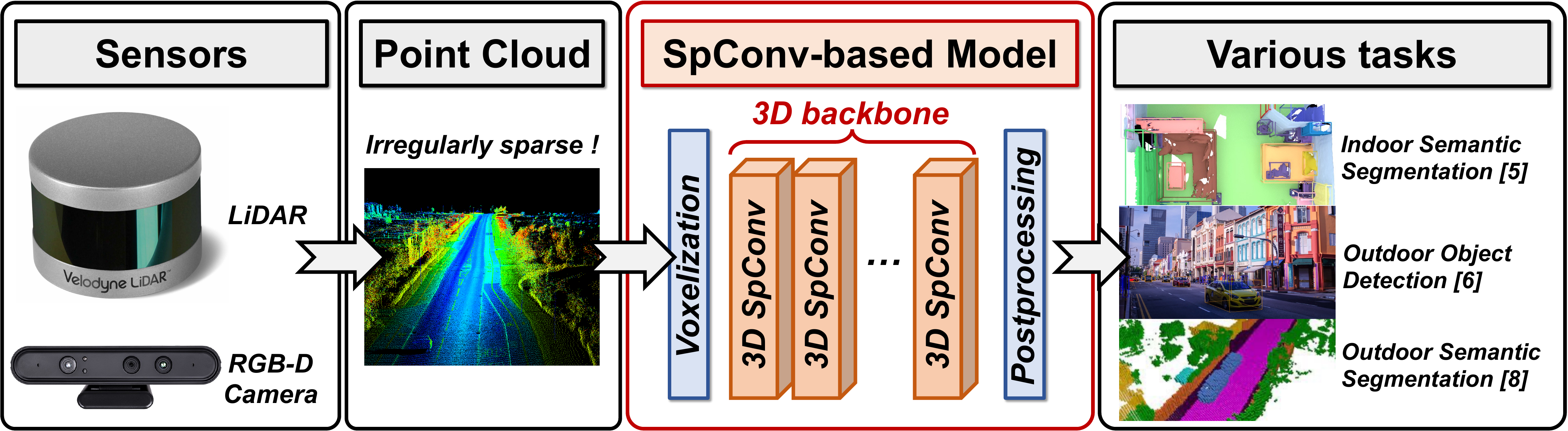}
    \caption{Point cloud neural networks with SpConv-based backbones \cite{Subm_CVPR19} has been widely used in various 3D tasks, such as indoor semantic segmentation\cite{MinkUNet_CVPR19}, outdoor object detection\cite{SECOND_Sensors18,CenterPoint_CVPR21} and semantic segmentation\cite{NASMIT_ECCV20}.}
    \label{fig:1_introduction}
\end{figure}

Furthermore, it also faces great challenges on the substantial increase in computation cost and parameter size of 3D convolution compared to traditional 2D CNNs.
The SpConv kernel is cube-shaped with an additional $z$-axis rather than a 2D square, enlarging the kernel size as well as Multiply-Accumulate (MAC) operations exponentially. 
Typical SpConv-based networks\cite{MinkUNet_CVPR19,NASMIT_ECCV20} incur 5-30$\times$ larger computation overhead than the widely-used ResNet\cite{ResNet_CVPR16}, which severely limits the performance of the previous works \cite{SpConvASIC_VLSI22,Tsinghua_ISSCC23}.

To tackle these dilemmas, we propose \emph{SpOctA}, a customized 3D sparse convolution accelerator designed for high-throughput and energy-efficient point cloud processing.
SpOctA leverages the fixed neighbor pattern in octree encoding to simplify the map search procedure, and fully utilizes the inherent sparsity and point distribution to improve the computation and memory efficiency.  
The key contributions are summarized as follows:
\begin{enumerate}
    \item The octree-encoding-based table-aided map search strategy is adopted to boost the parallelism potential of map search with algorithm-architecture co-optimization. 
    It introduces octree encoding to avert $O(n^2_{voxel})$-level traversing and improves hardware throughput with parallel query design.
    This approach attains 88.6-95.3\% reduction on search latency.
    \item A high-throughput sparsity-aware computing core is designed to eliminate the computing redundancy.
    By fully exploiting the inherent sparsity of each voxel, our method saves 44.4-79.1\% processing latency.
    \item A non-uniform caching memory management is proposed to apply different caching level for different parts of kernels. It achieves significant savings on on-chip storage and reduction in memory footprint.
\end{enumerate}

The SpOctA is implemented on a 40nm technology and fully evaluated with a set of widely-used SpConv-based models \cite{MinkUNet_CVPR19,SECOND_Sensors18} on 4 typical datasets\cite{Scannet_CVPR17,KITTI_CVPR12,SemanticKITTI_ICCV19,nuScenes_CVPR20}.
Compared with the prior state-of-the-art accelerators\cite{PointAcc_MICRO21,SpConvASIC_VLSI22,Tsinghua_ISSCC23}, our work achieves 1.1-6.9$\times$ speedup and 1.5-3.1$\times$ improvement on energy efficiency, supporting all kinds of typical SpConv operators.





\section{Preliminaries} \label{sec:II_Preliminary}

\subsection{3D Sparse Convolution in Point Cloud Networks}

In comparison with the conventional 2D dense convolution, 3D sparse convolution behaves differently in both tensor representation and computing rules\cite{Subm_CVPR19}.

\subsubsection{Sparse Tensor Representation}
2D images are always in dense format \cite{EIE_ISCA16}.
On the contrary, the voxelized point cloud data, noted as voxels, are usually represented as sparse tensors because of the extreme sparsity ($>$99\%), which follow the COO format \cite{MinkUNet_CVPR19} to store only the non-empty ones and their coordinates.
Each sparse tensor, denoted as $t_i$, contains two parts: voxel coordinates $\theta_i = (x_i, y_i, z_i)$ and associated feature map vector $f_i$.
We can represent the sparse tensor set of SpConv layer as:
\begin{equation}
    \mathcal{N}(t)= \{t_i = (\theta_i, f_i) | \theta_i\in \mathbb{Z}^3, f_i \in \mathbb{R}^{C_{in}}\}_{i=1\thicksim N_v} 
    \label{eq:2_sparse_tensor}
\end{equation}
where $N_v$ means the number of voxels in the current layer and $C_{in}$ is the number of input channels.

\subsubsection{SpConv Computing Rules}

\begin{figure}[t]
    \centering
    \includegraphics[width=\linewidth]{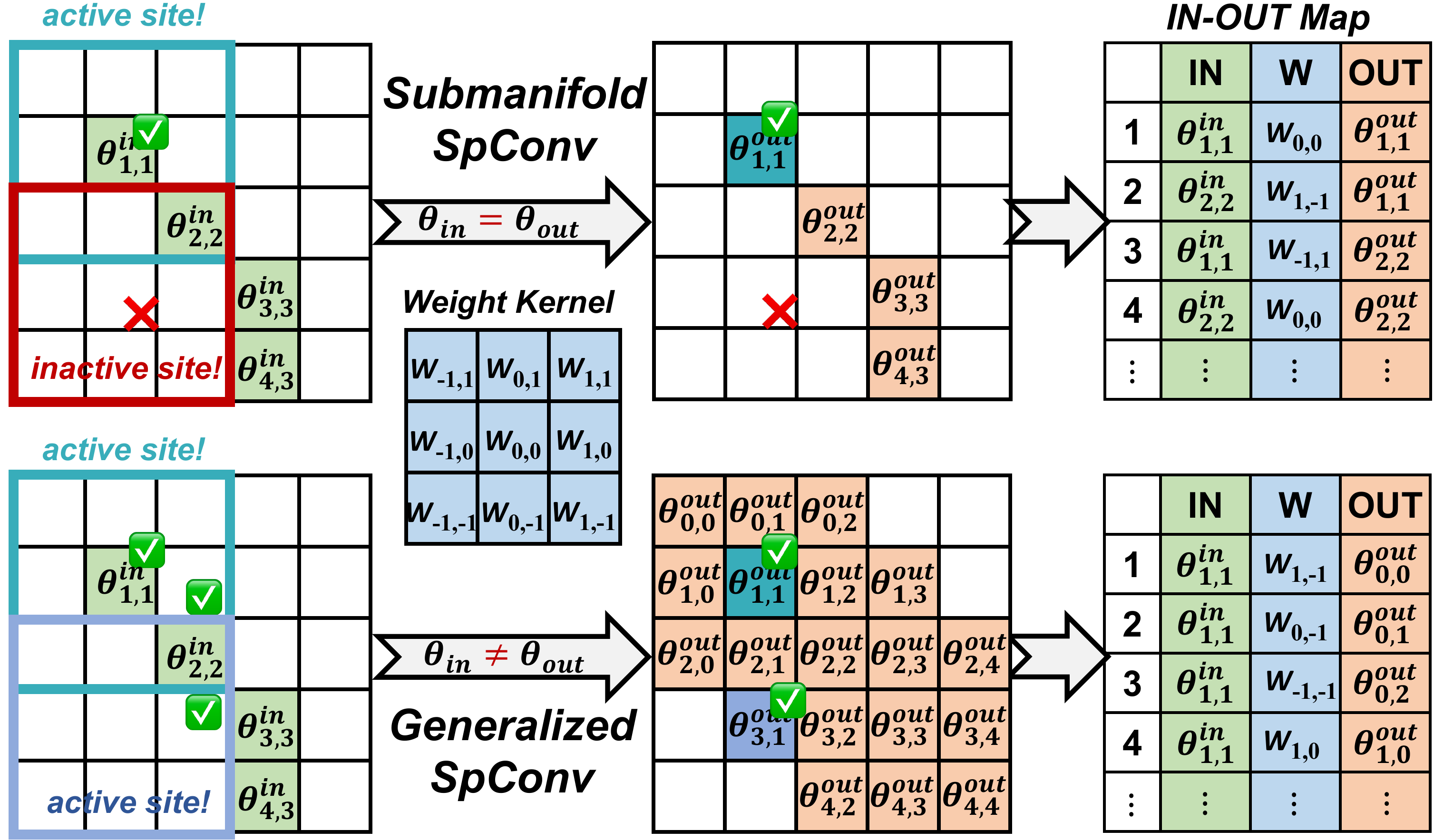}
    \caption{The special SpConv operator, submanifold SpConv\cite{Subm_CVPR19}, keeps the coordinates unaltered after the convolution, leading to a different map building rule from the generalized one.}
    \label{fig:2_spconv}
\end{figure}
For a SpConv layer of $K$-size kernel, the weight can be denoted as $W \in \mathbb{R}^{C_{out}\times C_{in} \times K^3}$ with kernel offset $\Delta^3_K$ from the kernel center.
For example, $\Delta^3_3 = \{\delta_x, \delta_y, \delta_z\}$, where $\delta_{x/y/z}=\{-1,0,1\}$.
Consequently, the SpConv operation can be represented as follows:
\begin{equation}
    t_i^{out} = (\mathcal{T}(\theta^{in}_i), \sum_{\delta \in \Delta^3_K}W_{\delta}f^{in}_j), \ \text{for} \ t_i^{in}, t_j^{in}\in \mathcal{N}(t^{in}) 
    \label{eq:2_spconv}
\end{equation}
where $\mathcal{T}$ is the transformation function of coordinates, and $t_j$ means all the active tensors covered by the sliding window.
Fig. \ref{fig:2_spconv} shows the processing examples of eq.(\ref{eq:2_spconv}).
Generalized SpConv executes the same as the conventional 2D convolution.
Meanwhile, SpConv also introduces a very different layer type: submanifold convolution \cite{Subm_CVPR19}, which executes only if the center site of sliding windows has non-empty voxels.
In this case, $\theta^{in}_i$ is equal to $\theta^{out}_i$.

Because of the special computing rules of submanifold SpConv, it needs map search to build the IN-OUT map from input feature maps (ifmap) at first to avoid the redundant operations\cite{Subm_CVPR19}.
It always traverses all the input possible voxels for each output voxel to find the candidates in the corresponding sliding window \cite{TorchSparse_MLSys22}.
The follow-up matrix multiplications in convolution are based on the map to gather the clustered values and scatter the partial sum to corresponding outputs.

\subsubsection{SpConv Layer Types}
Investigating on the representative benchmarks \cite{MinkUNet_CVPR19,SECOND_Sensors18,CenterPoint_CVPR21,NASMIT_ECCV20}, the most popular layer types can be summarized as follows:
\begin{itemize}
\item Submanifold SpConv: it is usually used as feature extraction layer. The popular kernel size is 3 and stride is 1. Here we note it as \emph{Subm3}.
\item Generalized SpConv: it is usually used as downsampling layer. The popular kernel sizes are 3 and 2, and the stride is always 2. Here we note them as \emph{Gconv3} and \emph{Gconv2} respectively.
\item Transposed SpConv: it is usually used to recover the original feature map with the same coordinates as that before the corresponding \emph{Gconv2}.
The popular kernel size is 2 and stride is always 2. Here we note them as \emph{Tconv2}.
\end{itemize}

\subsection{Motivation} \label{sec: II_B_motivation}

In contrast to the conventional 2D CNN, SpConv operators have raised variously new difficulties for the accelerations upon area-specific integrated circuits (ASIC).
Based on the sufficient investigations above, we summarize the following challenges:
\subsubsection{Large Memory Footprint and Latency of Map Search}
\begin{figure}[t]
    \centering
    \includegraphics[width=\linewidth]{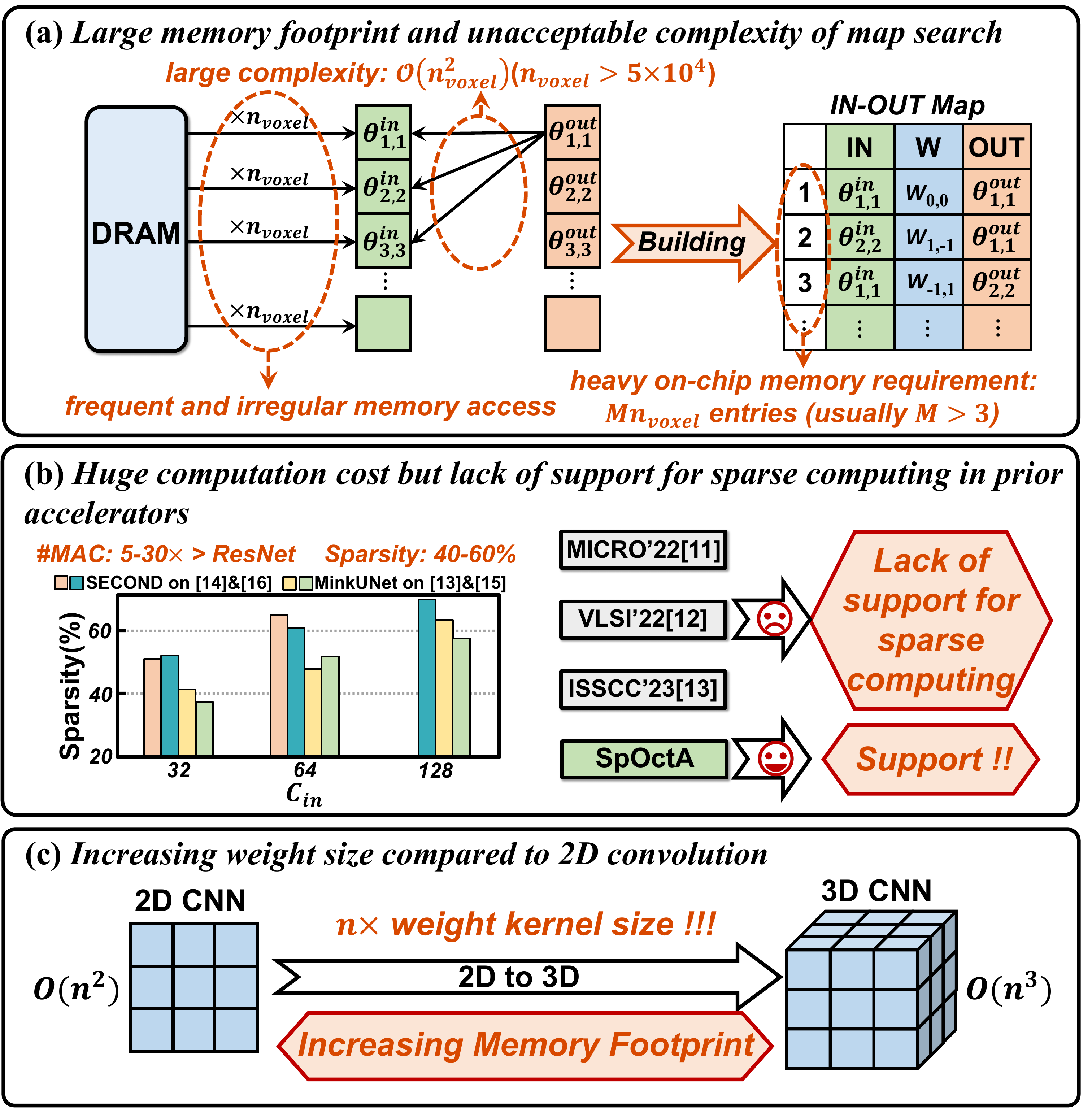}
    \caption{Three key challenges inspire our efforts of optimization}
    \label{fig:2_motivation}
\end{figure}

As described in Fig. \ref{fig:2_motivation}(a), map search is implemented via $\mathcal{O}(n_{voxel}^2)$-level voxel traversing, requiring huge data movement and frequent memory access explicitly.
It limits the hardware throughput due to the natural irregularity of point clouds.
Although state-of-the-art GPU-based implementations \cite{TorchSparse_MLSys22,MinkUNet_CVPR19,spconv2022} have attempted to address this issue by employing the hash-table-based solution, their approach cannot be deserialized efficiently and raises enormous on-chip memory requirement in ASICs.
\cite{PointAcc_MICRO21} unifies the map search to ranking-based paradigm for general point cloud applications so that it involves redundant logic design for SpConv operations.
\cite{SpConvASIC_VLSI22} reduces the search overhead by effective data skip techniques.
However, the acceleration is confined due to the huge processing latency of inevitably serial brute-force traversing.
\cite{Tsinghua_ISSCC23} proposes parallel priority-code-based search engine but it suffers from huge logic consumption of large multiplexer and priority coder.

\subsubsection{Huge Computation Cost but Lack of Support for Sparse Computing in Prior Accelerators}

3D SpConv neural networks consume 5-30$\times$ computation cost than traditional 2D CNNs\cite{ResNet_CVPR16}.
2D CNN accelerators\cite{EIE_ISCA16,SCNN_ISCA17,Eyerissv2_JETCAS19,STICKER_JSSC20,SNAP_JSSC21} commonly exploit the network sparsity\cite{EIE_ISCA16,SCNN_ISCA17} to attenuate the computational workload.
Similarly, our investigations, as illustrated in Fig. \ref{fig:2_motivation}(b), reveal that SpConv-based models\cite{MinkUNet_CVPR19,SECOND_Sensors18} exhibit 40-60\% sparsity in the typical benchmarks.
This provides great opportunities to reduce the excessive computation cost. 
Despite this potential, state-of-the-art SpConv accelerators \cite{PointAcc_MICRO21,SpConvASIC_VLSI22,Tsinghua_ISSCC23} do not capitalize on the inherent sparsity and, as a result, suffer from heavy computation overhead with limited acceleration performance.

\subsubsection{Tremendously Increasing Weight Size Compared to 2D Convolution}

As depicted in Fig. \ref{fig:2_motivation}(c), the weight kernel expands by a factor of $K$ because of an additional $z$-axis dimension introduced from 2D to 3D.
For a typical layer in MinkowskiUNet \cite{MinkUNet_CVPR19} with 8-bit quantization, the weight requires over 400KB of on-chip SRAM, which exceeds the memory capacity of most 2D CNN ASICs \cite{Eyerissv2_JETCAS19,STICKER_JSSC20,SNAP_JSSC21}.
While recent SpConv accelerators \cite{PointAcc_MICRO21,SpConvASIC_VLSI22,Tsinghua_ISSCC23} have employed various fetching or reusing methodologies to reduce weight memory consumption, they are still confronted with unavoidable challenges of tremendous DRAM and SRAM access.
Therefore, a better memory management strategy is imperative.

\section{Overviews of SpOctA Architecture} \label{sec:III_architecture}

\begin{figure}[t]
    \centering
    \includegraphics[width=\linewidth]{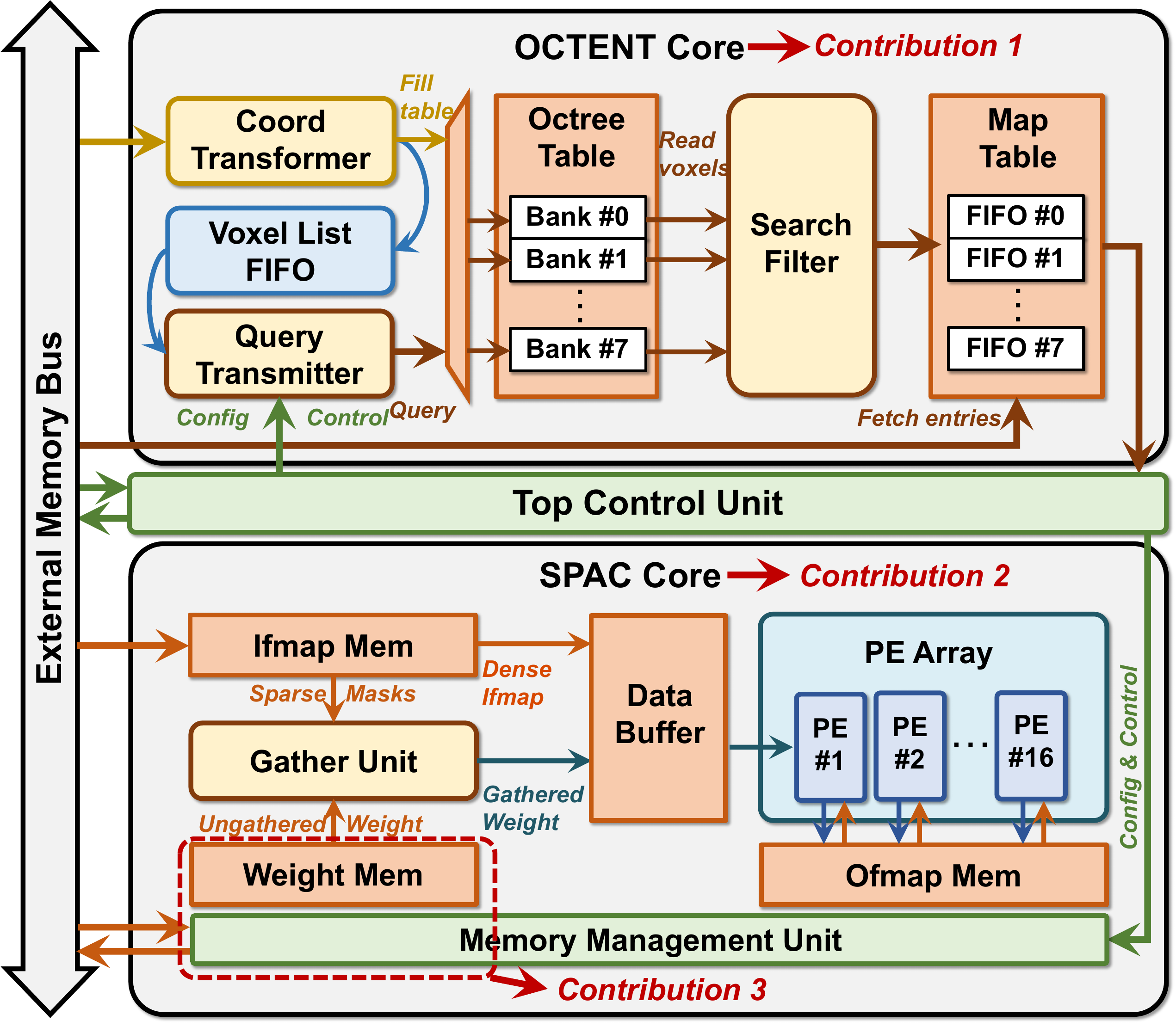}
    \caption{Overview of SpOctA architecture}
    \label{fig:3_architecture}
\end{figure}
To address the obstacles mentioned above, we present SpOctA, a high-throughput and energy-efficient architecture for SpConv processing.
As depicted in Fig. \ref{fig:3_architecture}, SpOctA consists of two cores: the \underline{OCT}ree-\underline{EN}coding-based \underline{T}able-aided map search (OCTENT) core, and the \underline{SP}arsity-\underline{A}ware \underline{C}omputing (SPAC) core.
Additionally, a top control unit connects the two cores by gathering the needed ifmaps and weights according to the IN-OUT maps and administrating the flow for execution.

SpOctA provides flexible support for various SpConv operators used in popular 3D point cloud models \cite{MinkUNet_CVPR19,SECOND_Sensors18,CenterPoint_CVPR21,NASMIT_ECCV20},
which attributes to the configurable search and computing cores.
Through parameter adjustment, the OCTENT core unifies search procedure for \emph{Subm3} and \emph{Gconv2}, or fetches map entries into Map Table directly for the other two operators.
In terms of the computing workload, the computing core operates with input/output stationary dataflow to support different reuse modes for various convolution types, reducing memory footprint significantly.
The contributions of architecture and dataflow design for these two cores marked in Fig. \ref{fig:3_architecture} are describled in the following sections.
\section{Efficient Octree-Encoding-Based Table-Aided Map Search Core}

This section proposes OCTENT, an algorithm-architecture optimization strategy for map search, which boosts the parallel potential in octree-encoding-based search method (Section \ref{Sec:4_octree_algo}) and achieves throughput improvement in circuit design (Section \ref{Sec:4_unit_design}\&\ref{sec:4_fine_grained_pipeline}).

\subsection{Algorithm Level: OCTENT Strategy for Parallel Search} \label{Sec:4_octree_algo}

\begin{figure*}[t]
    \centering
    \includegraphics[width=\linewidth]{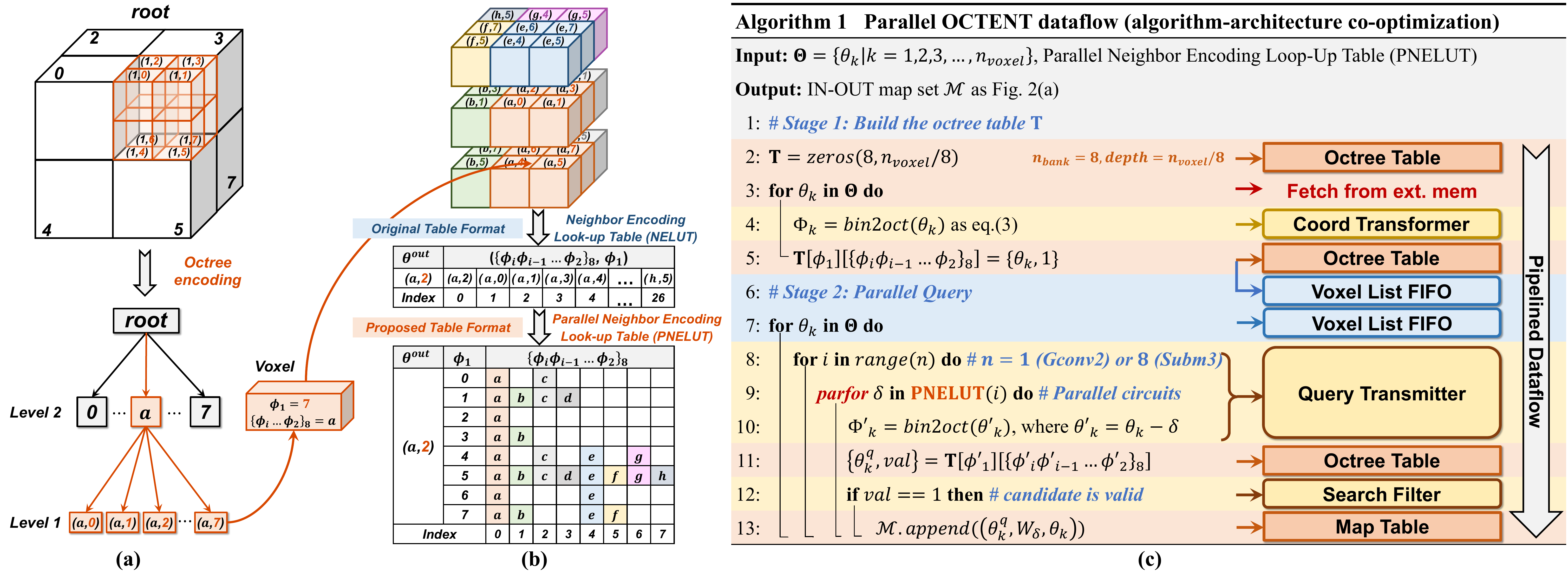}
    \caption{(a) Illustration of the octree-encoding rules for hierarchically cubical space.(b) An example of all the octree-encoded voxels in \emph{Subm3} sliding window. We reformat the NELUT to explore the parallelism potential of octree-encoding based search. (c) The detailed search procedure in OCTENT core.}
    \label{fig:4_octree_encoding}
\end{figure*}

\subsubsection{Octree Encoding}
Octree \cite{OctNet_CVPR17,Octree_Sensors18} is an effective tree-like data structure representing the sparse 3D space information with unbalanced encoding, which assigns the discrete voxels into leaf nodes by recursively partitioning the space into 8 cubes as depicted in Fig. \ref{fig:4_octree_encoding}(a).
For a 3D voxel with coordinate $\theta=(x,y,z)$, where $x=\{x_ix_{i-1}\cdots x_1\}_2, y=\{y_iy_{i-1}\cdots y_1\}_2,z=\{z_iz_{i-1}\cdots z_1\}_2$ in binary encoding, the octree code for the certain leaf node $\Phi$ can be written as:
\begin{equation}
        \Phi = (\phi_i,\cdots,\phi_1)= (\{z_iy_ix_i\}_2, \cdots, \{z_1y_1x_1\}_2)
    \label{eq:4_bin2oct}
\end{equation}
where $\phi_i$ means an octal value of the octree coordinate in the level $i$.
Fig. \ref{fig:4_octree_encoding}(b) illustrates an example of \emph{Subm3} sliding window under octree encoding and the octree codes for all the voxels are listed in Neighbor Encoding Look-up Table (NELUT).
The octree coordinates of the center is $(a,2)$ where $a = \{\phi_i\phi_{i-1}\cdots\phi_{2}\}_8$ and $\phi_1 = 2$, and the output coordinate $\theta^{out}$ is $(a,2)$ as well.
With the help of NELUT, building the IN-OUT map for this window is to traverse the NELUT, get the candidate codes and query it in the overall table for all possible voxels.
However, this procedure faces the same dilemma on deserialization as the hash-table-based method \cite{TorchSparse_MLSys22}, leading to large search latency.

\subsubsection{Parallel Search Strategy Based on Octree Encoding}
To address the issues, we propose the OCTENT strategy to boost the parallelism potential based on octree encoding.
At first, by extracting the lowest code $\phi_1$, we reformat the NELUT as Parallel NELUT (PNELUT) in Fig. \ref{fig:4_octree_encoding}(b). 
Candidates uniformly fall in each row so that constant parallelism occurs across different $\phi_1$.
Most importantly, using $\phi_1$ instead of $\phi_i$ could make the candidates be more evenly distributed in all rows, avoiding to limit the query throughput since concentrating in one row.
Based on the PNELUT, we construct the two-stage parallel search procedure as exhibited in Fig. \ref{fig:4_octree_encoding}(c).
At stage 1, the 2-dim octree table $\mathbf{T}$ is establish to accommodate all the non-empty voxels.
At stage 2, we traverse all the voxels (line 7), query the $\mathbf{T}$ based on PNELUT (line 8-11) and confirm whether the results are valid or not (line 12-13).
The parallel for-loop (\emph{parfor}) at line 9 sends the table queries concurrently towards different subarrays without collisions, and the iteration number $n$ is 1 for \emph{Gconv2} or 8 for \emph{Subm3}. 
Meanwhile, all the queries for arbitrary sliding windows can be generated based on the hardware-friendly PNELUT.
Because of the logic simplication compared to hash-table-based method\cite{TorchSparse_MLSys22}, it provides a feasible solution for parallel table-aided map search in terms of circuit specification.

\subsection{Architecture Level: OCTENT Core Implementation}\label{Sec:4_unit_design}

As we discussed above, the proposed OCTENT strategy can accomplish octonary-level parallelism from the \emph{parfor}.
We specialize a pipelined architecture for efficient implementation on ASIC platform, which is shown in Fig. \ref{fig:3_architecture}.
The detailed mapping strategy is also exhibited in Fig. \ref{fig:4_octree_encoding}(c) to describe the functions of each unit.
For the $1^{st}$ stage, the original coordinates $\theta_k$ are fed into Coord Transformer to get the octree codes $\Phi_k$ as table index.
After all the voxels are imported, the $2^{nd}$ stage begins.
Here we elaborate the query stage by introducing the key unit designs below.

\subsubsection{Octree Table}
The OCTENT approach is still unpractical due to the enormous size of table $\mathbf{T}$ when $n_{voxel}$ grows bigger than $5\times 10^4$ in most typical point cloud application\cite{Scannet_CVPR17,KITTI_CVPR12,SemanticKITTI_ICCV19,nuScenes_CVPR20}.
To reduce the resource consumption of $\mathbf{T}$, we adopt the block-wise processing strategy to restrict the searching space into small blocks (16$\times$16$\times$16 in this work).
It guarantees that all the voxels in one block can be cached on chip without extra off-chip DRAM access, which saves massive memory access latency as well as data movement energy. 

With the help of block partition, we are able to establish the octree table with sensibly sized on-chip SRAM shown in Fig. \ref{fig:4_timeline}(a), which contains 8 banks to maximize the parallelism.
Because the $\mathbf{T}$ is defined as a 2-dim array at line 2 in Algorithm 1, the subarray $\mathbf{T}[i]$ is accommodated in bank $i$, where $i$ represents the $\phi_1$ for the current voxel.
The corresponding element can also be read and written with address $\{\phi_i\phi_{i-1}\cdots\phi_2\}$.

\subsubsection{Query Transmitter}

Query Transmitter is designed to transmit 8 queries with different $\phi_1^{\prime}$ per cycle to fully utilize the bandwidth of multiple banks in Fig. \ref{fig:4_timeline}(a).
PNELUTs of different centers are built-in inside the Query Transmitter and it can be indexed by $\Phi_{center}$, $\phi_1$ and cycle counter \emph{cnt}.
For the vacancies in PNELUT, it just invalidates the control signals to mask the unnecessary queries.
This parallel transmitter can complete the search query in 8 cycles for \emph{Subm3} and only 1 cycle for \emph{Subm2}.
Once \emph{cnt} counts to zero, it enables the read signal of Voxel List FIFO and begins the round for the next voxel. 

\subsubsection{Search Filter and Map Table}
Search Filter aims to rectify query results for ordered Map Table writing.
It receives the query results with valid flags, filters out the invalid ones and gathers the valid candidates.
Then the Rectifier scatters an arbitrary number of valid maps based on the current FIFO status of Map Table.
For example, as shown in Fig. \ref{fig:4_timeline}(b), when two maps are valid and the writing pointer is 1, the rectifier redirects \emph{Map 0} to FIFO \#1 and \emph{Map 1} to FIFO \#2 and updates writing pointer to 3. 
As a result, the maps from the same sliding window will be vertically continuous among all FIFOs so that they are able to be read in sequence through polling.

\subsection{Fine-Grained Pipeline} \label{sec:4_fine_grained_pipeline}

\begin{figure}[t]
    \centering
    \includegraphics[width=\linewidth]{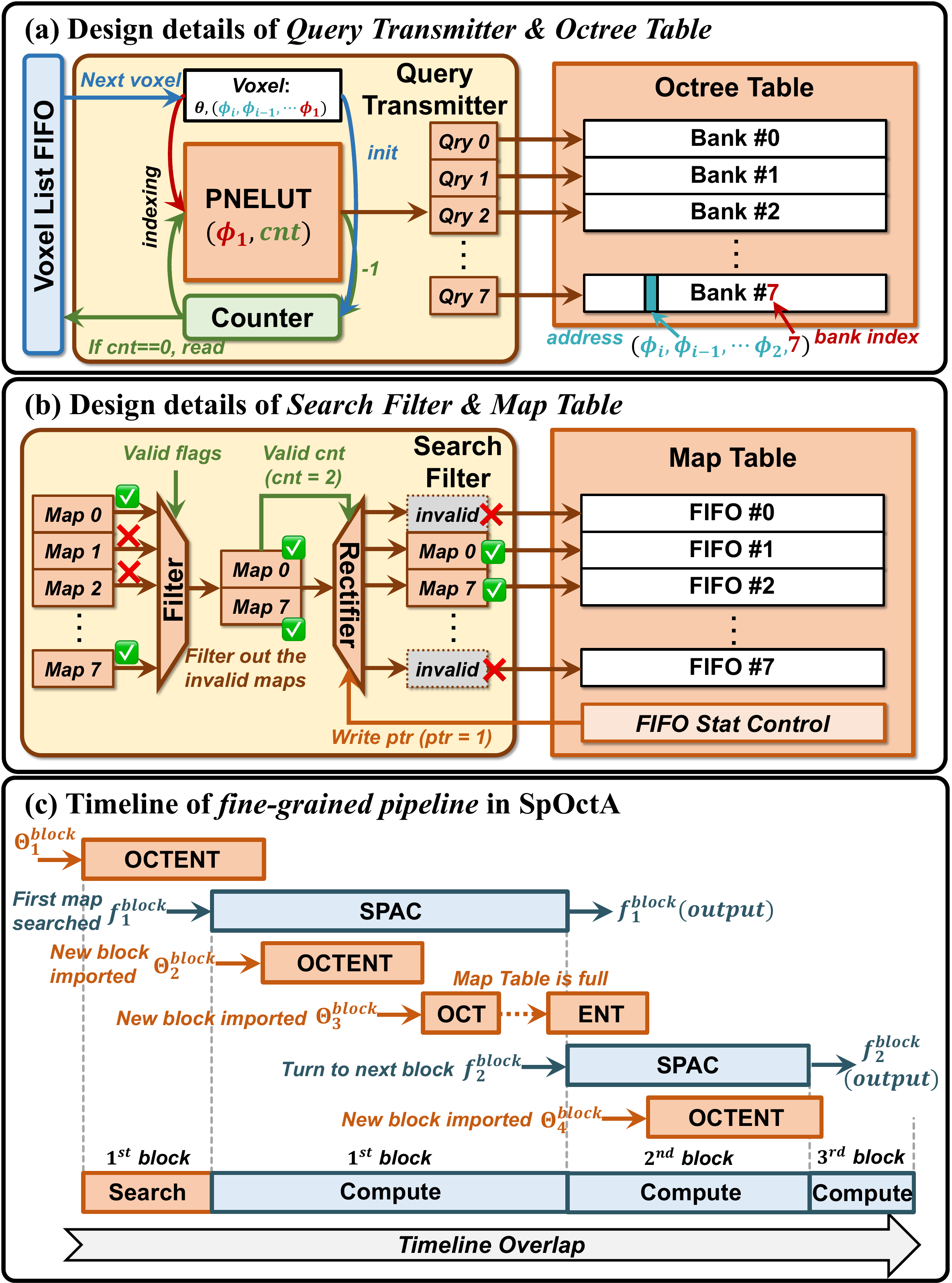}
    \caption{(a) \emph{Query Transmitter} sends the requests to \emph{Octree Table} in parallel. (b) \emph{Search Filter} filters out the invalid candidates and rectifies the map results to the right FIFOs in \emph{Map Table}. (c) The fine-grained pipeline between OCTENT and SPAC core covers the search latency effectively.}
    \label{fig:4_timeline}
\end{figure}

The work in \cite{SpConvASIC_VLSI22} constructs a coarse-grained pipeline by reusing the PE array during both map search and computing, which does not start the computing stage until all the IN-OUT maps have been searched.
This kind of strategy leads to a huge deceleration despite the significant area savings.
However, in our design, FIFO-based Map Table enables a fine-grained pipeline to accomplish throughput improvement compared to \cite{SpConvASIC_VLSI22}, and the detailed timeline is depicted in Fig. \ref{fig:4_timeline}(c) where $\Theta^{block}_k$ means the $\theta^{in}$ set of the $k^{th}$ block. 
Once a valid IN-OUT map in $1^{st}$ block is searched by OCTENT core, Top Control Unit reads it, fetches the $f_1^{block}$ and corresponding weights, and then the SPAC core begins the computing stage.
The OCTENT core keeps running block by block until the Map Table is full, and continues when next map is accessed.
It is obvious that the search and computing processing are overlapped by each other and most of the search latency is covered resulting from the fine-grained pipeline.

\subsection{Mapping Diverse Convolution Kernels on OCTENT Core}\label{Sec:4_pipeline}
Section \ref{Sec:4_unit_design} elaborates the map search procedure by taking \emph{Subm3} as an example.
Besides, OCTENT core also provides full supports of various SpConv operators, discussed as follows.

\subsubsection{\textit{Gconv2}}
\emph{Gconv2} is a key downsampling layer and its search space is restricted into an octree block with size of 8.
Consequently, PNELUT is simplified to 1 column and the counter in Query Transmitter is initialized to 1 per voxel for one-cycle query.

\subsubsection{\textit{Tconv2}}
The $\mathcal{M}_{Tconv2}$ is identical to $\mathcal{M}_{Gconv2}$ with $\theta^q_k/\theta_k$ switching.
To support it, we export $\mathcal{M}_{Gconv2}$ to external memory and reload it into Map Table while processing \emph{Tconv2} layers.

\subsubsection{\textit{Gconv3}}
Fig. \ref{fig:2_motivation}(a) shows the mapping details of \emph{Gconv3}, but the map search space will dilate to 5$\times$5$\times$5, causing huge increase on search latency and extra logic consumption.
In SpOctA, we decide to support \emph{Gconv3} by following an input stationary dataflow \cite{Tsinghua_ISSCC23} so that Map Table performs as an input buffer for original $\theta_k$ and the computing core reduces partial sums intelligently. 
\section{High Throughput Inherent Sparsity-Aware Computing Core}

\begin{figure}[t]
    \centering
    \includegraphics[width=\linewidth]{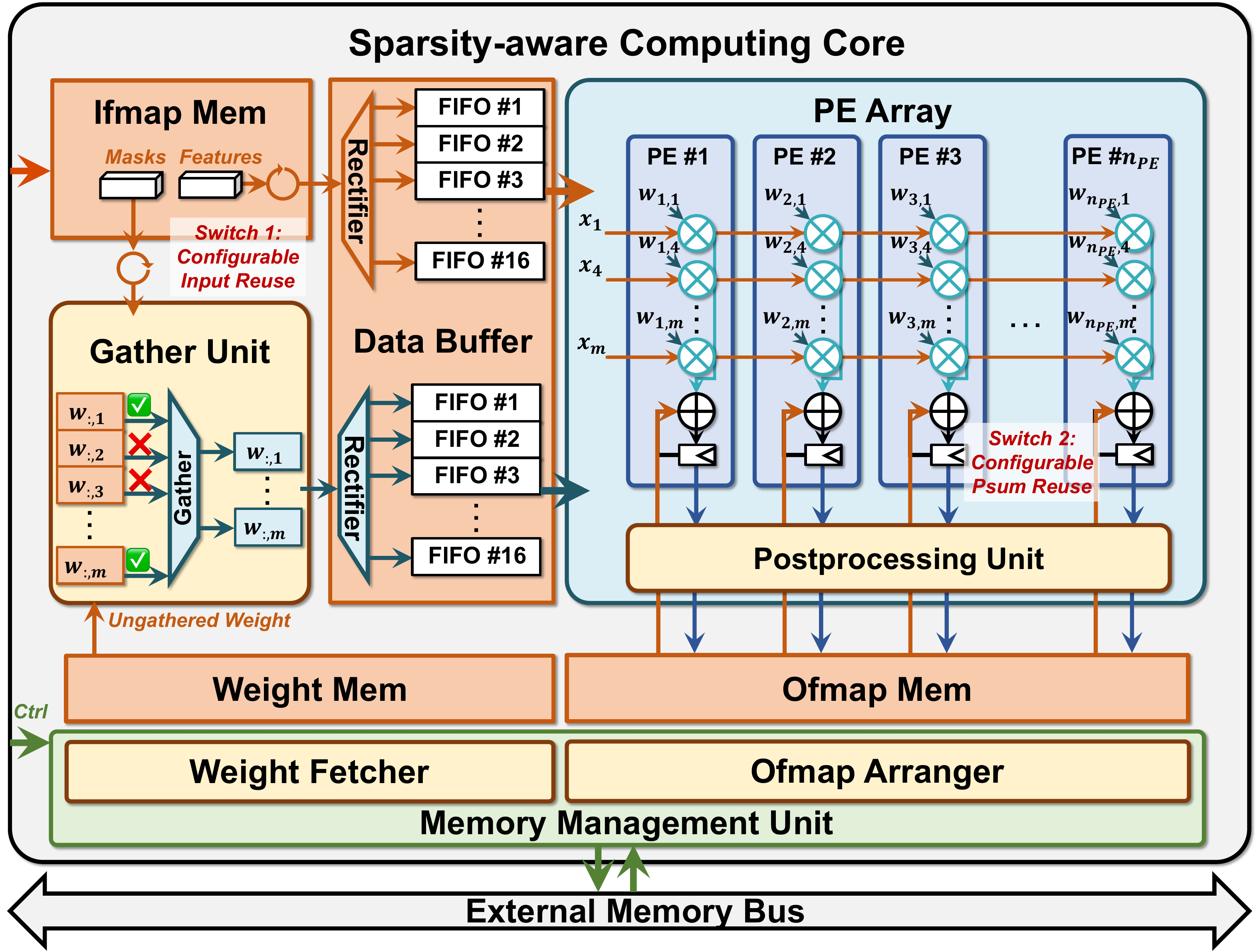}
    \caption{Architecture overview of the Sparsity-aware Computing Core}
    \label{fig:5_computing_core}
\end{figure}

The SPAC core is proposed to handle the computing workload in SpConv layers.
It deeply exploits the inherent sparsity to improve the computing efficiency (Section \ref{sec:5_sparse_computing}),
and manages the on-chip weight memory with non-uniform caching based on analysis of $W_{\delta}$'s distribution (Section \ref{sec:5_weight_memory}).

\subsection{Overview of the SPAC Core}

The SPAC core is composed of a PE Array for matrix multiplication, 3 parts of on-chip memory for ifmaps, weights and partial sums respectively, and other functional units, as illustrated in Fig. \ref{fig:5_computing_core}.
The PE Array is capable of $(16\times 16)\times(16\times 1)$ matrix-vector multiplication per cycle, parallelizing the computation on input channels and output channels and reusing ifmaps across PEs.
Diverse non-linear functions, like activation, batch normalization and quantization are supported in Postprocessing Unit.
Three parts of memory are organized to reduce the external memory access with a memory management unit, and they can be configured to switch between two dataflow mode: output stationary (\emph{Subm3}\&\emph{Gconv2}) and input stationary (\emph{Gconv3}\&\emph{Tconv2}).

\subsection{Sparse Computing Architecture} \label{sec:5_sparse_computing}

According to the analytic in Fig. \ref{fig:2_motivation}(b), ReLU function introduces 40-60\% inherent sparsity of ifmaps in average, resulting in huge redundancy.
Therefore, we design two peripheral units near PE Array to support sparse computing, Gather Unit and Ofmap Arranger.
Same as Search Filter in Fig. \ref{fig:4_timeline}(b), Gather Unit is built to gather the valid weights and ifmaps,
which strobes the unchosen weights with dynamic input masks and rectifies them for Data Buffer flexibly.
It guarantees fully exploitation of input sparsity to save huge computation cost as well as processing latency. 
As to output, Ofmap Arranger dynamically produces the masks of ofmaps per 16 output channels to match the parallelism of Gather Unit. 
Meanwhile, it can also judge if the accumulation for this sliding window is finished and generate writing request to external memory. 

\begin{figure}[t]
    \centering
    \includegraphics[width=\linewidth]{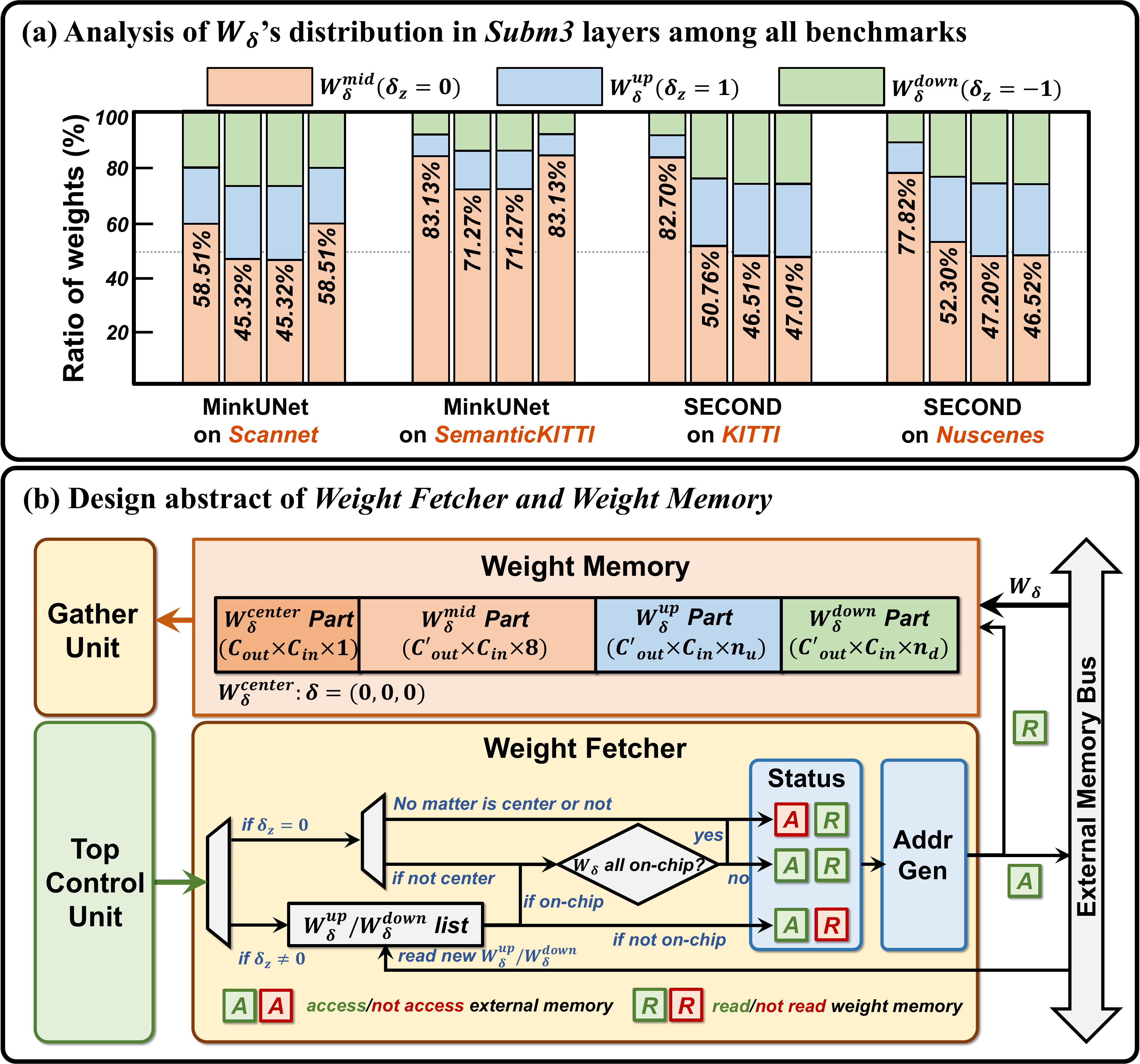}
    \caption{(a) Analysis of weight distribution of IN-OUT maps for different $\delta_z$ in four typical \emph{Subm3} layers among these well-known benchmarks. (b) Design abstracts of \emph{Weight Memory} and \emph{Weight Fetcher} with non-uniform caching strategy. }
    \label{fig:5_weight_management}
\end{figure}
\subsection{Non-Uniform Caching Strategy for Weight Memory Management} \label{sec:5_weight_memory}

Aiming to tackle the challenge of huge weight memory requirement described in Fig. \ref{fig:2_motivation}(c), we propose the non-uniform caching strategy for on-chip weight memory.

Point cloud sensors, especially LiDARs, always provide much higher horizontal resolution than vertical resolution because of laser scanning and object occlusion\cite{PCSurvey_PAMI21}.
Differences in resolution become more obvious after voxelization\cite{SpConvASIC_VLSI22}, which is reflected in vertical weight distribution.
Fig. \ref{fig:5_weight_management}(a) exhibits the ratios of three parts of weight kernels that are different in $\delta_z$.
It can be observed that the proportion of $W_{\delta}^{mid}$ reaches 45-83\% in all typical \emph{Subm3} layers, which is much larger than the other two parts.

Based on the analysis above, we develop a non-uniform weight memory organization for intelligent caching during processing.
As depicted in Fig. \ref{fig:5_weight_management}(b), the weight memory is divided into 4 partitions, and each partitions stores all or part of the corresponding values due to the configuration.
For example, $W_{\delta}^{center}$ partition contains all values of $W_{\delta}^{center}$ because it is certain to be involved for every sliding window.
We also tries to store all of $W_{\delta}^{mid}$ on chip but the partition is restricted to 32KB in some huge layers.
Instead, the memory storage of $W_{\delta}^{up}$ and $W_{\delta}^{down}$ partition can be tremendously saved with relatively small effect on the miss penalty.
Meanwhile, Weight Fetcher is designed to handle the cache-like logic workload.
It determines the map status based on the reserved value list and layer configuration.
The proposed strategy saves the memory consumption significantly.

\section{Evaluation}

\subsection{Experimental Setup}
\subsubsection{Benchmarks}

As Table \ref{tab:6_benchmark} exhibits, we choose 2 representative SpConv-based models, MinkowskiUNet\cite{MinkUNet_CVPR19} and SECOND\cite{SECOND_Sensors18}, for two promising applications: semantic segmentation and object detection.
Tests are on 4 widely-used datasets \cite{Scannet_CVPR17,SemanticKITTI_ICCV19,KITTI_CVPR12,nuScenes_CVPR20} sensed by RGB-D cameras and LiDAR sensors for indoor and outdoor scenes, evaluating our work thoroughly.

\subsubsection{Hardware Implementation}

We implement a prototype of SpOctA with Verilog and synthesis it with Synopsys Design Compiler under a 40nm CMOS technology.
A cycle-accurate simulator is developed to model the behavior of the architecture and to help the hardware verification.
Logic power consumption is estimated using PrimeTimePX by annotating switching activity from the benchmarks above, and SRAM area and energy are obtained with CACTI \cite{CACTI}.
For off-chip DRAM, we use a moderate DDR4 with 16GB/s bandwidth and the corresponding data access energy is 15pJ/b \cite{DRAM_Power}.

\subsubsection{Baselines}

We select various hardware platforms as the performance baselines, including powerful GPUs (NVIDIA 2080Ti \& 3090Ti GPU) and the state-of-the-art SpConv-based accelerators \cite{PointAcc_MICRO21,SpConvASIC_VLSI22,Tsinghua_ISSCC23}. 
As for GPUs, we reproduce the sparse convolution networks referring to the official implementions \cite{MinkUNet_CVPR19,NASMIT_ECCV20,mmdet3d2020,openpcdet2020} of the benchmarks to get the standard performance. 
For specialized accelerators, we evaluate the benchmarks on the SpOctA following the choices of them respectively for a fair comparison. 
All the networks are quantized to 8bits and retrained to recover the accuracy as the related works above.
Detailed results about throughput, latency, power consumption and energy efficiency are measured to exhibit the superiority of our work. 

\begin{table}[t]
    \centering
    \caption{Evaluation Benchmarks}
    \label{tab:6_benchmark}
    \renewcommand\arraystretch{1.4}
    \begin{tabular}{c|c|c|c}
        \toprule
        Application                                            & Dataset                      & Model                         & Notation \\
        \hline
        \multirow{2}{*}{\tabincell{c}{Semantic\\Segmentation}} & ScanNet\cite{Scannet_CVPR17} & MinkUNet(small)\cite{MinkUNet_CVPR19}& \emph{Seg(i)} \\
        \cline{2-4}
                                                               & SemanticKITTI\cite{SemanticKITTI_ICCV19} & MinkUNet(large)\cite{MinkUNet_CVPR19} & \emph{Seg(o)} \\
        \hline
        \multirow{2}{*}{\tabincell{c}{Object\\Detection}}      & KITTI\cite{KITTI_CVPR12} & SECOND(small)\cite{SECOND_Sensors18}& \emph{Det(k)} \\
        \cline{2-4}
                                                               & nuScenes\cite{nuScenes_CVPR20} & SECOND(large)\cite{SECOND_Sensors18} & \emph{Det(n)} \\
        \bottomrule
    \end{tabular}
\end{table}

\subsection{Evaluation Results of Optimization Tactics} \label{sec:6_self_evaluation}
\begin{figure}[t]
    \centering
    \includegraphics[width=\linewidth]{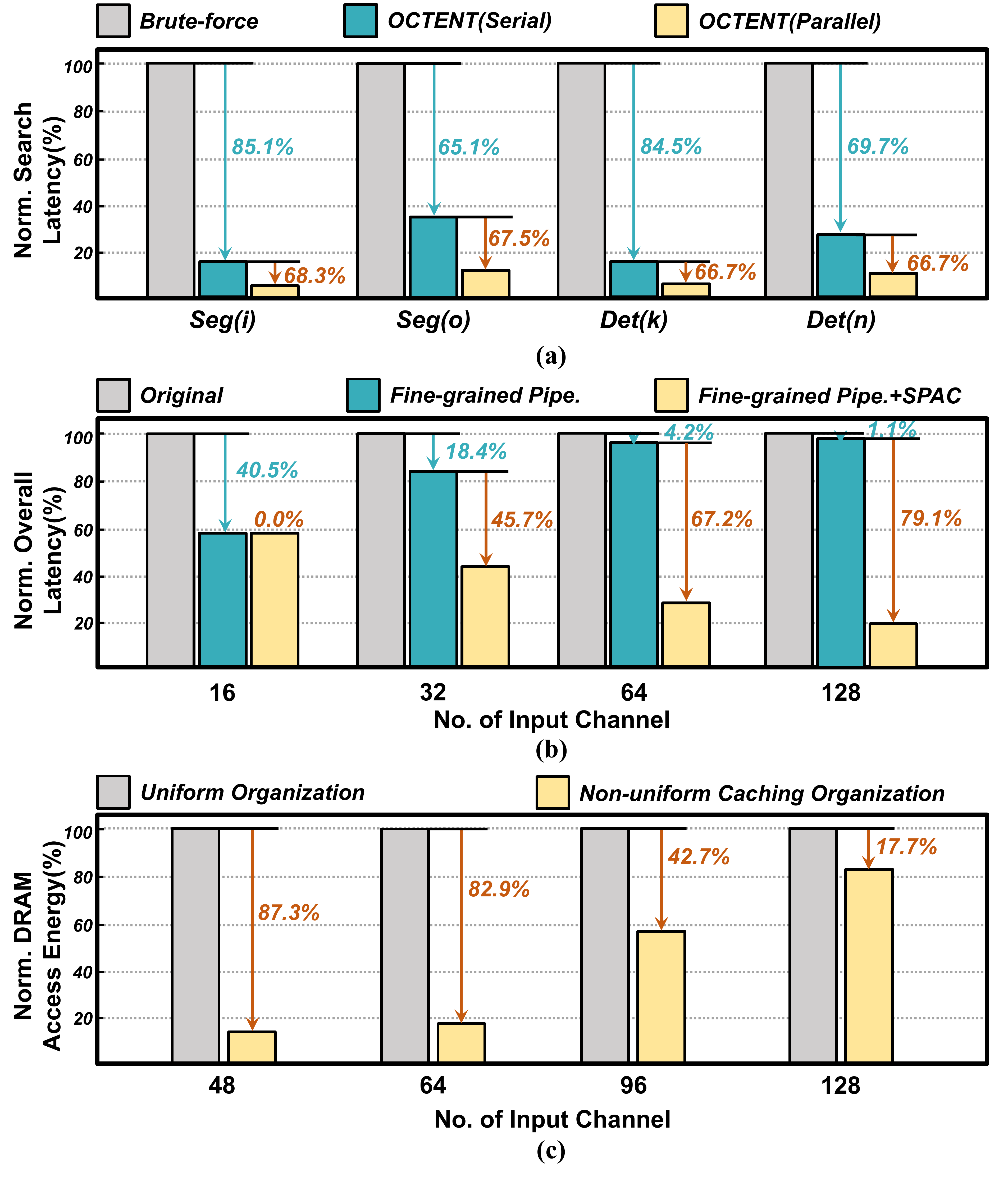}
    \caption{(a) illustrates the average latency reduction of each benchmark with our OCTENT algorithm-hardware co-optimization. (b) shows the normalized reduction on overall latency benefitting from fine-grained pipeline and SPAC. (c) exhibits savings of DRAM access energy because of non-uniform caching management for weight memory. Analysis in (b) and (c) is based on all the layers with typical input channel numbers.}
    \label{fig:6_self_evaluation}
\end{figure}

To provide a comprehensive demonstration on our optimization tactics, we conduct experiments based on the benchmarks in Table \ref{tab:6_benchmark} to evaluate the improvements in detail.
Results are depicted in Fig. \ref{fig:6_self_evaluation} and discussed as follows.

\subsubsection{Parallel OCTENT Algorithm-Architecture Co-Design}

Fig. \ref{fig:6_self_evaluation}(a) shows the evaluation results of search latency reduction by proposed OCTENT strategy.
Results are averaged by all \emph{Subm3} and \emph{Gconv2} layers.
It is obvious that more than 65\% search latency can be saved by involving serial OCTENT algorithm, which does not unroll the for-loop at line 9 in Fig. \ref{fig:4_octree_encoding}(c).
With the query loop deserialized by OCTENT algorithm and architecture, huge hardware throughput is improved explicitly. 
Experiments demonstrates that 66.7-68.3\% latency reduction is further achieved, resulting in 8.8-21.2$\times$ speedup on map search altogether.

\subsubsection{Fine-Grained Pipeline and Sparsity-Aware Computing Paradigm}

Fig. \ref{fig:6_self_evaluation}(b) shows the evaluation results of the latency reduction associated with fine-grained pipeline (Section \ref{Sec:4_unit_design}) and sparsity-aware computing paradigm (Section \ref{sec:5_sparse_computing}).
Since the speedup tremendously varies across the input channel numbers, we average the results by layers with 4 different $C_{in}$s in all benchmarks.
The acceleration from fine-grained pipeline achieves up to 1.68$\times$ at $C_{in}$=16 on account of part of search latency covered by the computing effectively.
Although its effect decays when computing workload dominates the overall latency as growth of $C_{in}$ (like 128), the sparsity-aware computing contributes significantly.
It is proven that nearly 80\% runtime saving can be obtained by SPAC.

\subsubsection{Non-Uniform Caching Strategy}

Fig. \ref{fig:6_self_evaluation}(c) shows the evaluation results of external memory access for weights, categorized by $C_{in}$ as well.
The on-chip memory is set large enough to hold all the weights of layers with $C_{in}\leq32$.
Up to 87.3\% DRAM access energy can be saved at $C_{in}=48$ resulting from great miss rate decreasing of $W_{\delta}^{center}$ \& $W_{\delta}^{middle}$, which also reduces significantly large on-chip memory accesses as well.
Although the saving ratio decays along with $C_{in}$ increasing due to limited memory capacity, involving the non-uniform strategy still saves external data movement by more than 42\% at $C_{in}=96$ and 17\% at $C_{in}=128$ on average, which is definitely a tremendous improvement on memory efficiency.

\subsection{Comparisons with the State-of-the-Art Works}

\begin{figure*}[t]
    \centering
    \includegraphics[width=\linewidth]{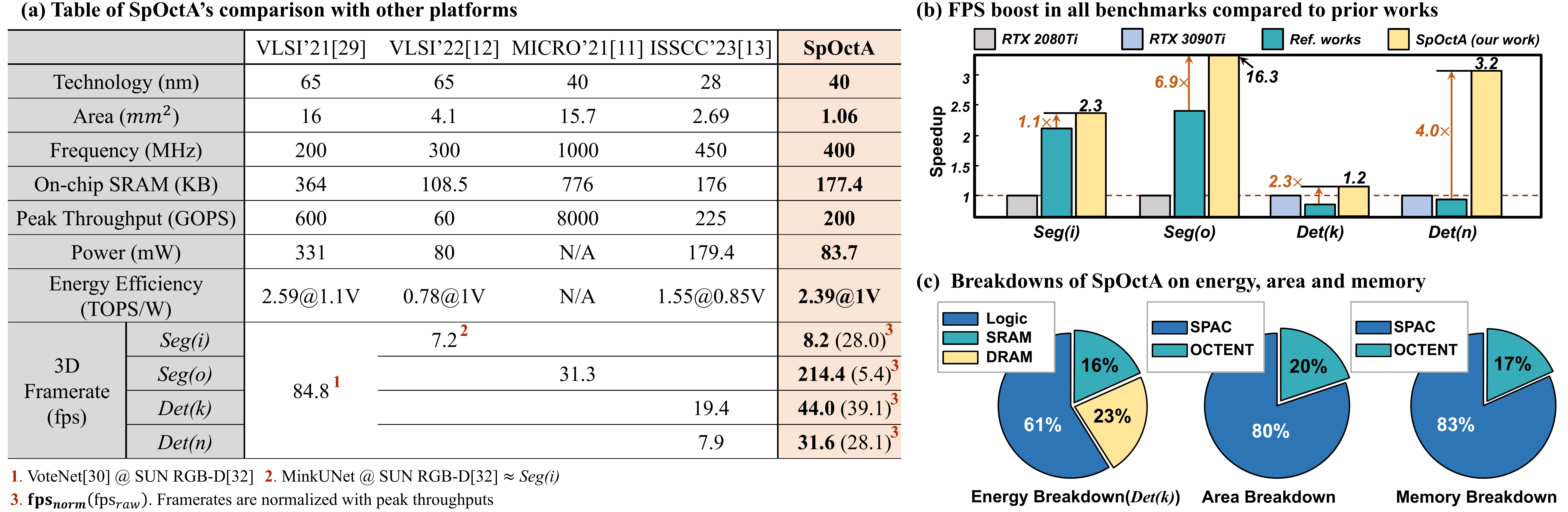}
    \caption{(a) Detailed comparison table of SpOctA and related accelerators. (b) Normalized speedup by SpOctA compared to prior works and powerful NVIDIA GPUs. (c) Overall implementation breakdown of SpOctA on energy, area and memory consumption.}
    \label{fig:6_comparison}
\end{figure*}

Fig. \ref{fig:6_comparison}(a) presents the detailed implementation results of SpOctA.
SpOctA achieves 200GOPS at 400MHz while dissipating 83.7mW at 1V and occupying 1.06mm$^2$.
Fig. \ref{fig:6_comparison}(c) breaks down the hardware consumption.
It can be noticed that DRAM accesses consume only 23\% energy benefiting from the effective management strategies, and OCTENT core only covers 20\% area overhead with 17\% on-chip SRAM.

Comparisons with state-of-the-art point-cloud accelerators are depicted in Fig. \ref{fig:6_comparison}(a) as well, where \cite{PNNPU_VLSI21} focuses on PointNet-based algorithm\cite{VoteNet_2019ICCV} and \cite{PointAcc_MICRO21,SpConvASIC_VLSI22,Tsinghua_ISSCC23} are for SpConv-based networks.
Although our benchmark \emph{Seg(i)} applies 2.5$\times$ larger dataset (50k points per frame in ScanNet\cite{Scannet_CVPR17}) and 5.8$\times$ deeper network, comparable energy efficiency and framerates (84fps normalized to 600GOPS) with \cite{PNNPU_VLSI21} are also attained on indoor segmentation task.
Meanwhile, SpOctA provides complete support of promissing sparse convolution operators instead of PointNet-based network \cite{PointNet++_NeurIPS17}, which is inefficient and impractical for large-scale point cloud processing, such as urban autonomous driving scenes\cite{KITTI_CVPR12,nuScenes_CVPR20}.

Compared to the related SpConv-based works \cite{PointAcc_MICRO21,SpConvASIC_VLSI22,Tsinghua_ISSCC23}, our work achieves 1.5-3.1$\times$ energy efficiency improvement with 1.1-6.9$\times$ acceleration because of tactics evaluated in Section \ref{sec:6_self_evaluation}.
The detailed speedup is exhibited in Fig. \ref{fig:6_comparison}(b), which is normalized with the peak throughputs for a fair comparison.
\cite{SpConvASIC_VLSI22} reuses the PE array for map search to save area and power overhead, but it limits the possibility of pipeline processing. 
Fine-grained pipeline between search and computing makes our work outperform \cite{SpConvASIC_VLSI22} on \emph{Seg(i)} benchmark despite testing on a heavier workload \cite{Scannet_CVPR17} than SUN RGB-D\cite{SUNRGBD_CVPR15}.
\cite{PointAcc_MICRO21} unifies all kinds of point cloud convolution into ranking-based computation to accomplish excellent compatibility, involving redundant operations for SpConv inevitably.
SpOctA focuses on SpConv operators and simplifies the map search through OCTENT methodology so that 6.9$\times$ speedup is achieved compared to \cite{PointAcc_MICRO21} on \emph{Seg(o)} test. 
For object detection benchmarks \emph{Det(k)}\&\emph{Det(n)}, the work in \cite{Tsinghua_ISSCC23} adopts parallel search engine with smart multi-core scheduling to improve the throughput significantly, but the framerates that are less than 20fps are still insufficient at urban autonomous driving scenes.
Our work introduces sparsity-aware computing strategy to reduce massive computation cost and processing latency, boosting the framerates by 2.3-4.0$\times$ for a more satisfactory real-time performance.

\section{Conclusion}

In this manuscript, we present SpOctA, a SpConv accelerator to achieve high-speed and energy-efficient processing on extensive point cloud applications.
SpOctA presents octree-encoding-based search optimization to enable parallel map search in ASIC platform, boosting the hardware throughput with low logic overhead significantly. 
To handle the heavy computation workload of SpConv, our work exploits the inherent sparsity of each voxel to eliminate the redundancy in matrix multiplication.
SpOctA also achieves great savings on the consumption of external memory access through non-uniform caching strategy which leverages the irregular resolution at different axises.
Combining these innovations, SpOctA delivers 1.1-6.9$\times$ speedup with 1.5-3.1$\times$ higher energy efficiency over state-of-the-art accelerators on plentiful benchmarks, providing effective support of various SpConv kernels.

\bibliographystyle{IEEEtran}
\bibliography{IEEEabrv,Ref}

\vfill

\end{document}